\documentclass{jpp}
\usepackage{graphicx}

\usepackage[utf8]{inputenc}
\usepackage[T1]{fontenc}
\usepackage{amsmath}
\usepackage{natbib}
\usepackage{subcaption}
\usepackage{bm}

\shorttitle{RL-Based Plasma Control on EXL-50U}
\shortauthor{P. Guo, Z. Chen, J. Chen, X. Wang, G. Shi, S. Ding, Y. Zhang, L. Xing, T. Liu}

\title{State-Space model-enabled Reinforcement Learning for magnetic configuration control on EXL-50U}

\author{
Pei Guo\aff{1},
Zhengyuan Chen\aff{2}\corresp{\email{czyneu@163.com}},
Jianguo Chen\aff{2},
Xuanhe Wang\aff{2},
Guoyang Shi\aff{2},
Siqi Ding\aff{2},
Yapeng Zhang\aff{2},
Lei Xing\aff{2},
Yong Liu\aff{2},
Xiang Gu\aff{2},
Tiantian Sun\aff{2},
Xiuchun Lun\aff{2},
Jia Li\aff{2},
Zhengxiong Wang\aff{1}\corresp{\email{zxwang@dlut.edu.cn}},
Huasheng Xie\aff{2},
Hanyue Zhao\aff{2},
Yuejiang Shi\aff{2},
Xianming Song\aff{2},
Tianyuan Liu\aff{2},
EXL-50U Team\aff{2}
}

\affiliation{
\aff{1}Key Laboratory of Materials Modification by Beams of the Ministry of Education, School of Physics, Dalian University of Technology, Dalian 116024, China
\aff{2}ENN Science and Technology Development Company, Langfang, China
}

\begin{document}

\maketitle

\begin{abstract}
Accurate feedback control of the plasma current ($I_p$) and centroid position $(R_c,Z_c)$ is essential for the stable operation of spherical torus (ST) plasmas. Conventional proportional–integral–derivative (PID) controllers require extensive manual tuning and struggle with the fast, strongly coupled dynamics that arise as plasma performance improves. Reinforcement learning (RL) has recently emerged as a promising alternative to such complex magnetic control problems, yet its practical deployment on ST devices remains challenging. This paper presents a practical RL controller for the EXL-50U ST, trained within a rigid RZIP state-space model (SSM) that enables efficient offline policy learning. A lightweight plasma position re-constructor is developed to estimate $(R_c,Z_c)$ from magnetic probe signals within the real-time control cycle. The trained policy is seamlessly deployed on the EXL-50U plasma control system, achieving stable regulation of $I_p$ and $(R_c,Z_c)$ and sustaining discharges up to 650 ms under RL control. These results demonstrate the feasibility and practical potential of model-informed RL for magnetic configuration control in ST devices, offering a promising direction beyond conventional PID-based schemes.
\end{abstract}


\section{Introduction}
Reliable control of the plasma current $I_p$ and centroid position $(R_c, Z_c)$ is a critical requirement for spherical torus discharge operation. The low aspect ratio ($A \leq 1.5$), high $\beta_N$, and large elongation \citep{pengST,ST-Features} of ST plasmas induce fast, nonlinearly coupled magnetohydrodynamic (MHD) instabilities \citep{wanglong}, posing significant challenges for control system design. The standard approach employs poloidal field (PF) and central solenoid (CS) coils as actuators, regulating $I_p$ and $(R_c,Z_c)$ by adjusting coil currents and voltages to generate a controllable poloidal field. 

\begin{figure}
    \centering
    \includegraphics[width=0.6\linewidth]{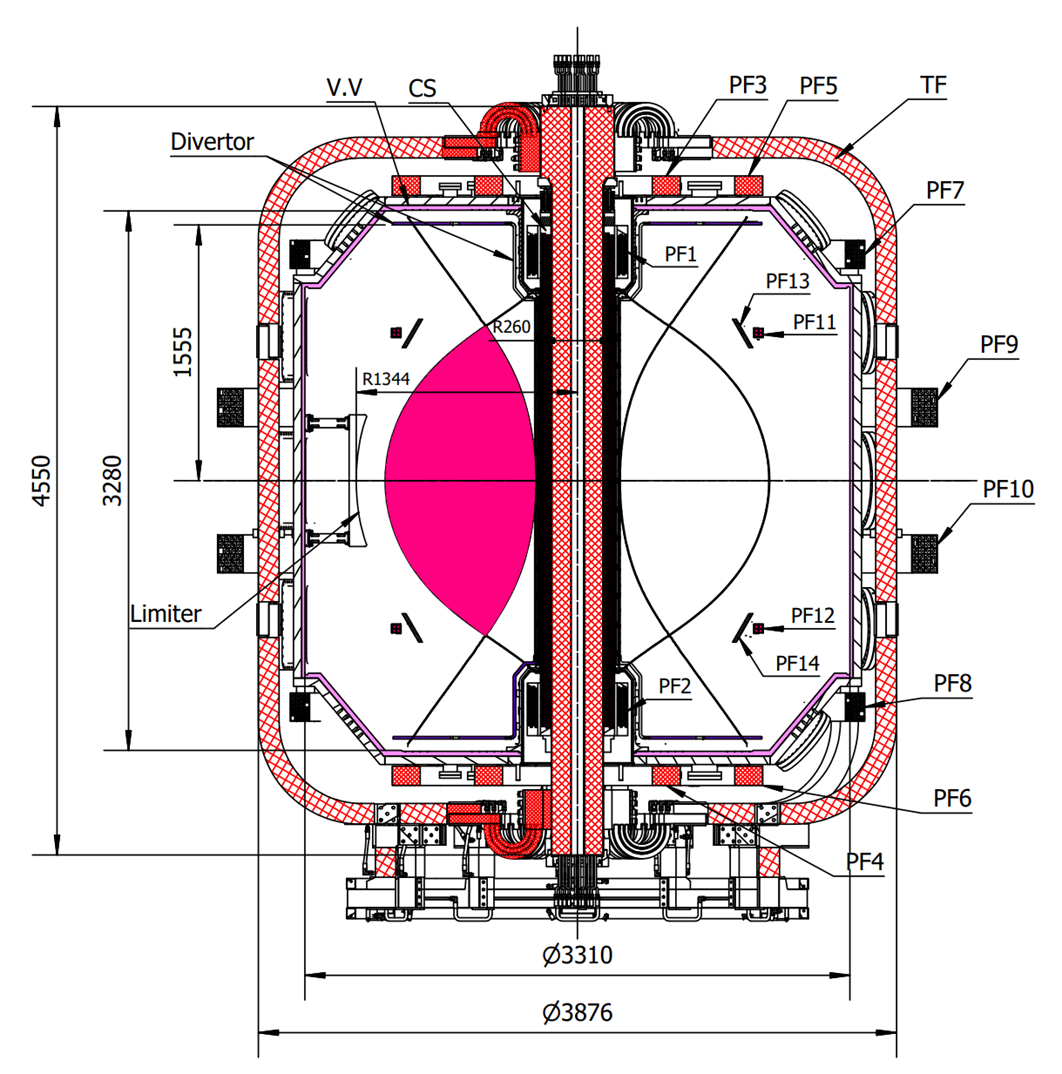}
    \caption{Poloidal cross-section of the EXL-50U device, where CS denotes Central Solenoid coils, V.V denotes Vacuum Vessel, TF denotes Toroidal Field coils, and PF denotes Poloidal Field coils.}
    \label{fig:EXL50U}
\end{figure}

The EXL-50U ST \citep{ennroadmap,syj,1MA}, which is dedicated to exploring high-performance plasma scenarios and validating key technologies for proton–boron fusion energy capture, thus demands stable and robust feedback control of $I_p$ and $(R_c,Z_c)$ to sustain discharge experiments. EXL-50U features an integrated center-stack magnet configuration, consisting of toroidal field (TF) coils, a CS coil, and 14 PF coils, shown as figure~\ref{fig:EXL50U}. The TF coils provide the main toroidal confinement field, while the CS coil acts as the ohmic heating coil to initiate and sustain the plasma current. PF1–PF6 serve as divertor configuration coils, in which PF1–PF2 can assist in plasma position control and PF3–PF6 are mainly used for plasma elongation; PF7–PF10 are responsible for global plasma equilibrium and displacement control, with PF7–PF8 dominating vertical position regulation and PF9–PF10 specializing in horizontal position control; PF11–PF12 are in-vessel fast control coils, also known as vertical stability (VS) coils, used to suppress vertical displacement events (VDEs), and PF13–PF14 act as passive coils for additional vertical stability. This functionally partitioned coil system enables flexible and precise control of the plasma configuration, position, and vertical stability for stable operation.

Currently, EXL-50U adopts incremental single-input single-output (SISO) PID feedback loops combined with pre-programmed feedforward waveforms to regulate the $I_p$ and $(R_c, Z_c)$ \citep{syj}, as shown in figure~\ref{fig:framework}. As a classic closed-loop control algorithm, PID adjusts the control output through proportional, integral and derivative terms based on the real-time error between the target and measured values, and its incremental implementation is well-suited for the real-time digital PCS of EXL-50U. Although this method can achieve basic stabilization, it suffers from several limitations: heavy manual tuning, limited scalability for multi-variable control objectives, and the need for repeated on-site parameter adjustments when operating conditions or control targets change \citep{plasma-ctrl}. To achieve more adaptive, robust, and high-performance plasma control, researchers have increasingly introduced advanced control strategies from other fields.

Machine learning has emerged as a transformative tool in fusion research, with extensive applications in tokamak discharge prediction \citep{Wan_2021,Wan_2022}, disruption prevention \citep{Guo_2021,Yang_2020}, and modeling of plasma transport and instability \citep{LI_2021,Li_2022,Zhao_2022,xu2024neural}. As a key subfield focused on sequential decision-making, reinforcement learning excels at multi-input multi-output (MIMO) control and is well suited for the complex, nonlinear, and multiscale dynamics of tokamak discharges \citep{wiering2012reinforcement,HL3}. It has shown great potential for plasma current control, configuration design, scenario tracking and instability avoidance \citep{deepmind,kstar,d3dRL,EASTRL,Tartaglione_2025,seo2024avoiding}. However, a major challenge hindering the practical deployment of RL on ST devices lies in the training environment. High-fidelity physics-based simulators that solve the Grad–Shafranov (GS) equation \citep{gs} provide accurate dynamics but are computationally prohibitive for the extensive iterative training required by RL. Conversely, purely data-driven simulators demand large amounts of experimental data and often fail to generalize across different tokamaks. To overcome this challenge, we adopt a computationally efficient rigid RZIP state-space model as the RL training environment, which strikes a favorable balance between physical fidelity and computational speed. 

The deployment of the RL policy trained with the RZIP model requires real-time state observations, including the plasma current $I_p$ and the centroid position $(R_c,Z_c)$. On EXL-50U, $I_p$ can be directly measured by a Rogowski coil with high accuracy and low latency, satisfying the real-time requirement. In contrast, the centroid position $(R_c,Z_c)$ cannot be measured directly and must be reconstructed from magnetic probe signals. However, the existing equilibrium reconstruction tool (EFIT \citep{EFIT}) on EXL-50U is designed for offline analysis and cannot operate at the required 1 kHz real-time control frequency. To enable online RL deployment, we adopt a lightweight RZEstimator \citep{RZob} that rapidly reconstructs $(R_c,Z_c)$ from a limited set of magnetic probe signals, providing sufficiently accurate estimates within the 1 kHz control cycle.

The main contributions of this study are threefold:
\begin{itemize}
    \item \textbf{SSM-based RL training environment}: A physics-constrained linearized RZIP model designed for efficient policy optimization of $I_p$ and $(R_c,Z_c)$ control on EXL-50U.
    \item \textbf{Low-latency real-time position estimation}: A lightweight \textbf{RZEstimator} that provides high-precision $(R_c,Z_c)$ estimates within the millisecond control cycle of EXL-50U.
    \item \textbf{On-device experimental validation}: Reproducible online deployment with smooth PID-RL control handover, achieving up to 650 ms of stable control of $I_p$ and $R_c$.
\end{itemize}

The remainder of this paper is organized as follows. Section~2 introduces the proposed methodology, including the SSM training environment, RL algorithm design, and on-device deployment pipeline. Section~3 presents comprehensive experimental results and performance analysis on the EXL-50U ST. Section~4 discusses the key findings, limitations, and potential future research directions.

\section{Methods}    
\subsection{Framework Overview}

\begin{figure}
    \centering
    \includegraphics[width=0.8\linewidth]{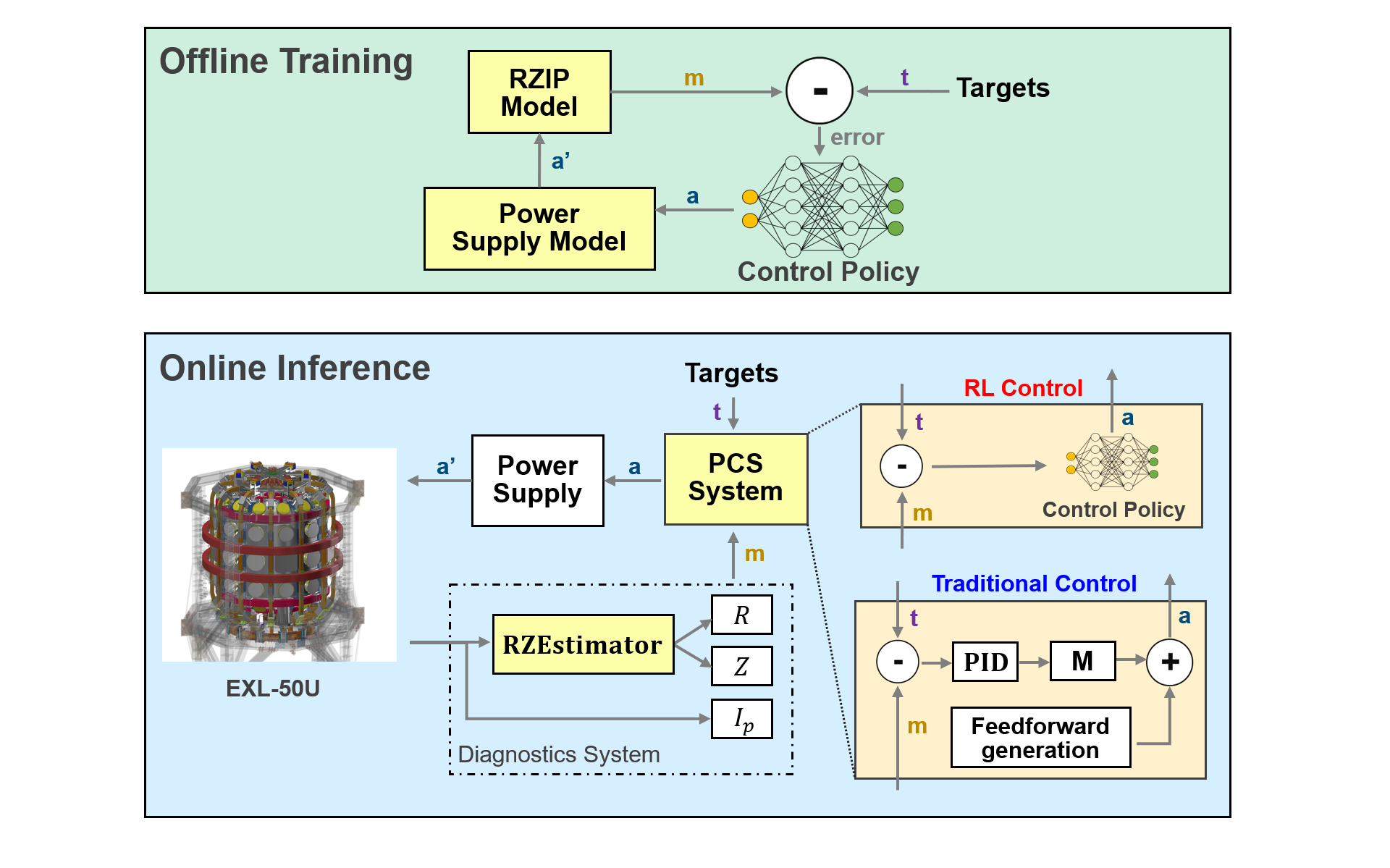}
    \caption{Overall workflow of the RL-based plasma magnetic control system for EXL-50U. The workflow consists of three key stages: physics-informed model development, offline RL policy training, and real-time on-device deployment.}
    \label{fig:framework}
\end{figure}

This work establishes a RL-based closed-loop control framework for EXL-50U plasma magnetic regulation, integrating an offline policy training stage and an online real-time inference stage, as illustrated in figure~\ref{fig:framework}. In the offline stage, a fast physics-informed simulation environment, composed of the rigid RZIP SSM and a power supply model, serves as the training platform, where the RL control policy is optimized using the Proximal Policy Optimization (PPO) algorithm. In the online stage, the RL policy is deployed on the EXL-50U plasma control system (PCS). The PCS acquires real-time state observations from the diagnostic system, along with target plasma parameters$(I_p,R_c,Z_c)$. The PCS supports two parallel control paths: the conventional PID feedback plus pre-programmed feedforward scheme, and the proposed RL-based controller. In RL-controlled discharges, the RL policy directly generates voltage commands for the coils based on the received observations and targets. These commands are then output to the coil power supplies, closing the control loop on the EXL-50U device.

\subsection{Physics-Informed Simulation Environment}

A practical RL training environment must capture the dominant plasma-coil coupling dynamics relevant to the control objectives while remaining computationally efficient for large-scale policy optimization. Although GS-based equilibrium models \citep{gs} offer high fidelity, they are computationally prohibitive for extensive iterative simulations in RL training and unnecessary for regulating only the global $I_p$ and $(R_c,Z_c)$. Instead, we adopt the rigid RZIP model, which strikes a favorable balance between physical fidelity and computational efficiency, making it well-suited as the core simulation environment for policy optimization.

\subsubsection{Rigid RZIP Model}
\label{sec:rzip_model}

The RZIP model is a canonical linearized rigid plasma response model widely used for the design and analysis of axisymmetric magnetic control in tokamaks \citep{walker2006valid,rzip-tcv}. It is derived under the core assumption that the plasma behaves as a rigid current-carrying conductor, allowing only global radial and vertical displacements of the plasma centroid while maintaining a fixed internal current profile. By neglecting plasma internal profile evolution and shape deformation, this model simplifies the complex nonlinear MHD equilibrium governed by the G–S equation into a computationally tractable linear framework, well-suited for plasma control system synthesis. 

The RZIP model is formulated based on circuit equations derived from Faraday's law, describing the electromagnetic coupling between structured conductors ($s$: active coils and passive conducting structures) and the plasma ($p$). The circuit equations for structured conductors and plasma are consolidated as:
\begin{align}
    \frac{d(\mathsfbi{M}_{ss} \boldsymbol{I}_s)}{dt}+\frac{d(\boldsymbol{M}_{sp} I_p)}{dt} + \mathsfbi{R}_{s} \boldsymbol{I}_s  &= \boldsymbol{V}_s \\
    \frac{d(\boldsymbol{M}_{ps} \boldsymbol{I}_s)}{dt} + L_p \frac{dI_p}{dt} + R_p I_p &= V_{\text{n.o.}}
\end{align}
where:
\begin{itemize}
    \item $\boldsymbol{I_s}$: Current vector of structured conductors;
    \item $I_p$: Total plasma current;
    \item $\boldsymbol{V_s}$: Voltage vector applied to coil conductors;
    \item $V_{\text{n.o.}}$: Effective voltage applied to drive plasma current by noninductive sources;
    \item $\mathsfbi{M_{ss}}$: Mutual inductance matrix between structured conductors;
    \item $\boldsymbol{M_{sp}} / \boldsymbol{M_{ps}}$: Mutual inductance vectors between structured conductors and plasma ($\boldsymbol{M_{sp}} = \boldsymbol{M_{ps}}^\text{T}$);
    \item $\mathsfbi{R_s} / R_p$: Resistance matrices of structured conductors/plasma;
    \item $L_p$: Plasma self-inductance.
\end{itemize}

This system is transformed into the standard linear state-space representation:
$$ \dot{\boldsymbol{x}} = \mathsfbi{A} \boldsymbol{x} + \mathsfbi{B} \boldsymbol{u} $$
$$ \boldsymbol{y} = \mathsfbi{C} \boldsymbol{x} + \mathsfbi{D} \boldsymbol{u} $$
where:
\begin{itemize}
    \item $\boldsymbol{x}$: State vector comprising all coil and plasma currents ($\boldsymbol{x} = \boldsymbol{I} = [\boldsymbol{I}_s^T,I_p]^T$);
    \item $\boldsymbol{u}$: Input vector corresponding to the coil voltage vector ($\boldsymbol{u} = [\boldsymbol{V}_s^T,0]^T$);
    \item $\boldsymbol{y}$: Output vector including $(R_c,Z_c)$;
    \item $\mathsfbi{A}$: System matrix characterizing the intrinsic dynamics of the system;
    \item $\mathsfbi{B}$: Input matrix describing how control inputs affect the state $\boldsymbol{x}$;
    \item $\mathsfbi{C}$: Output matrix mapping system states $\boldsymbol{x}$ to controlled variables $\boldsymbol{y}$;
    \item $\mathsfbi{D}$: Direct transmission matrix, set to zero as standard for tokamak systems.
\end{itemize}

This state-space framework enables the implementation of advanced control strategies (PID, optimal control, $H_\infty$ synthesis \citep{HUMPHREYS2008193,948481}) for robust plasma magnetic control design. With the rigid RZIP state-space model established, a practical simulation environment still requires a power supply model. In the actual system, the voltage commands issued by the PCS are not perfectly reproduced at the coil terminals; there exists a non-negligible discrepancy between the commanded and the actual applied voltages. To simulate the closed-loop dynamics accurately, this discrepancy must be captured. Various methods exist for constructing a power supply model, including transfer function approximations, Simulink-based block diagrams, and data-driven approaches (e.g., neural networks) \citep{PSM1,PSM2,PSM3}. Given that our primary objective is to validate the feasibility of the RL framework rather than to develop a high-fidelity power supply model, we adopt the simplest linear approximation:
$$ \boldsymbol{u}_{\text{sim}} = \alpha \cdot\boldsymbol{u}_{\text{com}} + \beta $$
where $\boldsymbol{u}_{\text{sim}}$ represents the simulated actual coil voltage vector, $\boldsymbol{u}_{\text{com}}$ denotes the PCS voltage command vector, $\alpha$ is the constant scaling coefficient, and $\beta$ is the constant offset coefficient. Identical values of $\alpha$ and $\beta$ are adopted for vertically symmetric coil pairs. The calibrated simulated voltage vector $\boldsymbol{u}_{\text{sim}}$ serves as the formal input voltage for the rigid RZIP model.

\subsubsection{Model Validation}
\label{sec:model_validation}

\begin{figure}
    \centering
    \includegraphics[width=0.75\linewidth]{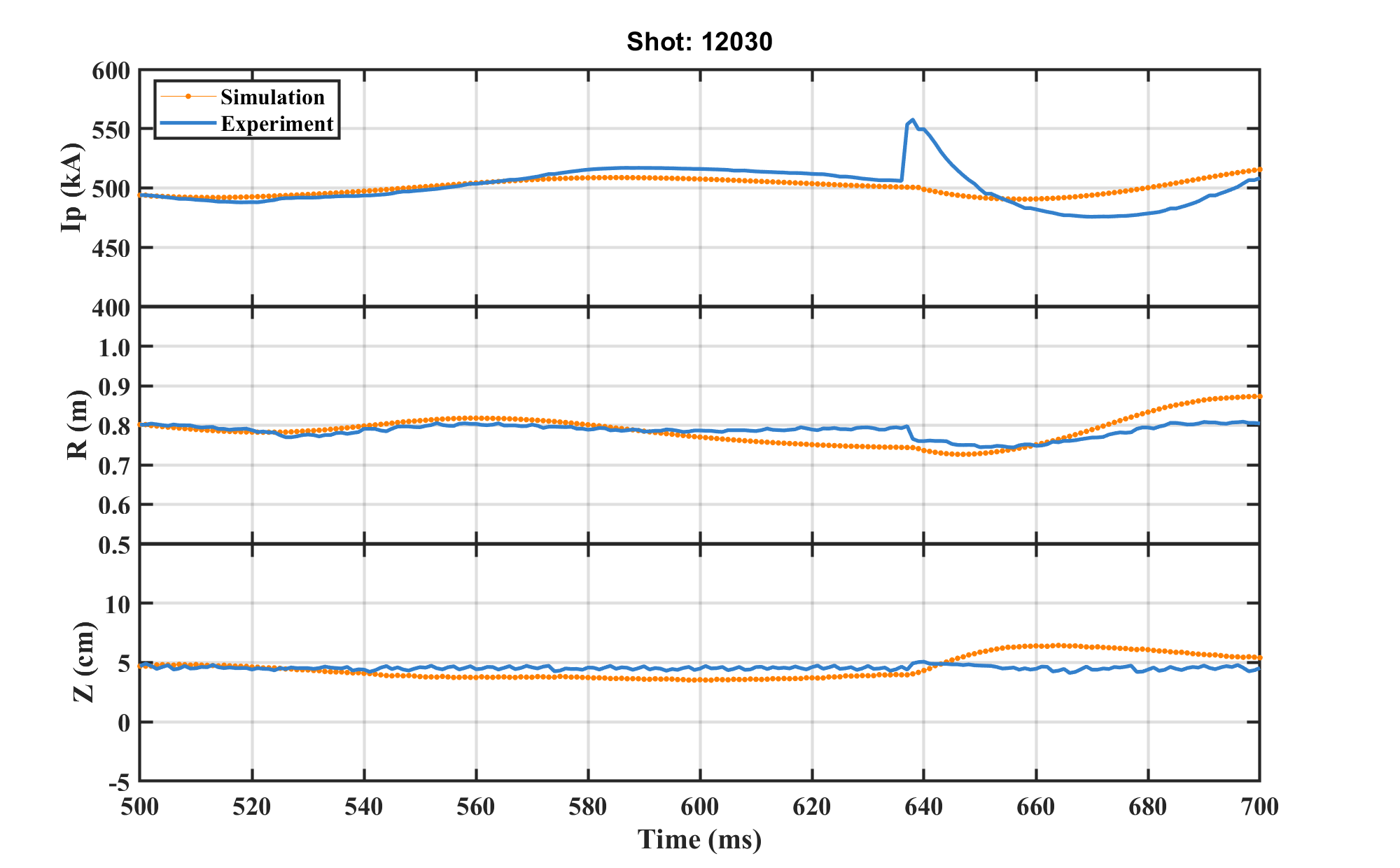}
    \caption{Comparison of simulated (RZIP model) and experimental plasma dynamics for EXL-50U reference discharge \#12030, including $I_p$, $R_c$ and $Z_c$ trajectories.}
    \label{fig:RZIP_evolution}
\end{figure}

\begin{figure}
    \centering
    \includegraphics[width=0.9\linewidth]{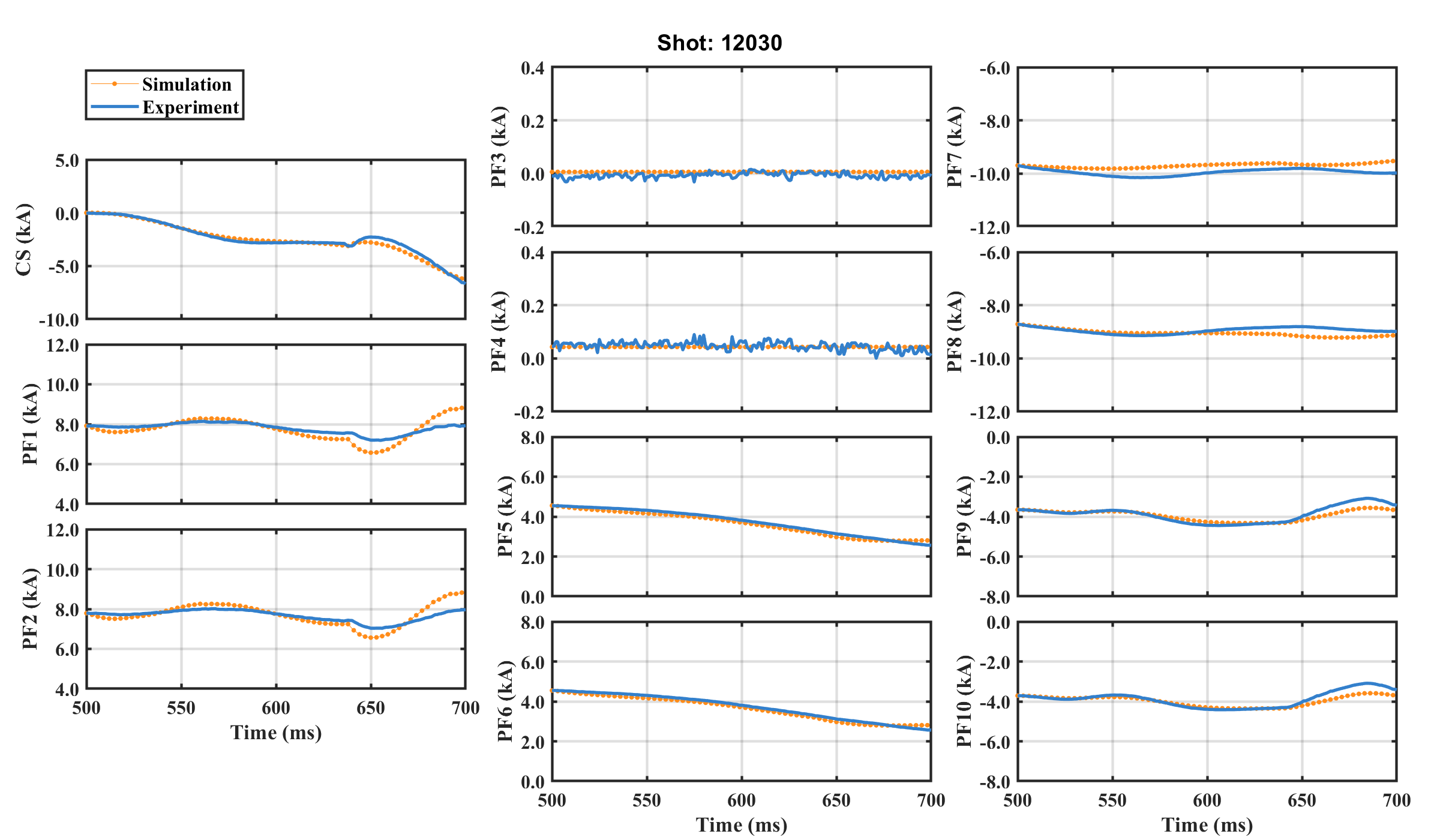}
    \caption{Comparison of simulated (RZIP model) and experimental CS and PF coil current waveforms for EXL-50U reference discharge \#12030.}
    \label{fig:coil_currents_evolution}
\end{figure}

The rigid RZIP SSM was validated against experimental data from EXL-50U reference discharge \#12030. The model was initialized to the 500\,ms limiter equilibrium profile of this discharge (achieved via pre-programmed waveforms) and driven by experimentally measured coil voltage waveforms starting from 500\,ms. Figure~\ref{fig:RZIP_evolution} compares the simulated and experimental trajectories of plasma current $I_p$ and centroid position $(R_c,Z_c)$, while figure~\ref{fig:coil_currents_evolution} shows the corresponding comparison of CS and PF1–PF10 coil currents.

Quantitative comparisons across both plasma dynamics and coil currents demonstrate that the rigid RZIP model effectively captures the dominant dynamic features of EXL-50U limiter discharges. The simulated $I_p$ and $(R_c,Z_c)$ closely follow the experimental trajectories, confirming the model’s accuracy in reproducing global $I_p$ and $R_c$ evolution. Nevertheless, non-negligible discrepancies remain: the simulated $Z_c$ shows limited dynamic accuracy, which is attributed to the simplified representation of the fast active vertical stabilization system in the model. Additionally, while the overall trends of all coil currents are well reproduced, quantitative deviations are observed for some PF coils.

\subsection{RL Control Framework}
Conventional SISO PID controllers require extensive manual tuning and make limited use of inter-coil electromagnetic coupling in multi-variable control. To address this limitation, we cast the simultaneous regulation of $I_p$ and $(R_c,Z_c)$ as a RL control problem. The environment dynamics are described by the rigid RZIP SSM (Section~\ref{sec:rzip_model}), while the agent outputs coil voltage commands subject to practical hardware constraints. The control policy is trained offline using the PPO algorithm \citep{PPO}, validated in closed-loop simulations, and then deployed online to the EXL-50U PCS without additional on-site fine-tuning.
\subsubsection{Reinforcement Learning Background}
\label{sec:rl_background}
Reinforcement learning provides a general framework for learning closed-loop control policies through repeated interaction between an agent and an environment \citep{sutton1998reinforcement}. In a standard Markov decision process (MDP), the environment is characterized by a state space $\mathcal{S}$, an action space $\mathcal{A}$, a transition law $P(s_{t+1}|s_t,a_t)$, and a scalar reward signal $r_t=R(s_t,a_t)$. At each discrete time step, the agent selects an action according to a policy $\pi_\theta(a_t|s_t)$ parameterized by $\theta$, the environment evolves to the next state, and the agent receives a reward evaluating the quality of the transition. The policy optimization objective is to maximize the expected discounted return
\begin{equation}
J(\theta)=\mathbb{E}_{\pi_\theta}\left[\sum_{t=0}^{T-1}\gamma^t r_t\right],
\end{equation}
where $\gamma\in[0,1]$ is the discount factor and $T$ is the episode horizon. For continuous-control problems, the policy is commonly represented by a neural network, enabling a direct mapping from measured system quantities to actuator commands.

In this study, the policy is optimized using PPO \citep{PPO}, an on-policy actor--critic algorithm designed to improve training stability by limiting how much the policy is allowed to change after each batch of simulated interactions. PPO updates the policy through a clipped surrogate objective, which discourages overly large policy-ratio changes while still permitting multiple gradient updates on collected rollout data. This property is useful for plasma magnetic control because unstable exploratory updates can quickly drive simulated discharges outside the valid operating region, whereas overly conservative updates would make training inefficient. PPO therefore provides a practical compromise between implementation simplicity, stable learning, and compatibility with continuous voltage-command actions in the RZIP training environment.

This formulation can be specialized to plasma magnetic control by identifying the plasma-coil system as the environment, the coil voltage commands as actions, and the tracking quality of plasma quantities as the reward. The central design task is therefore to construct a simulator in which trial-and-error learning is feasible, and to define a reward that reflects the desired magnetic-control objectives. Recent tokamak studies have shown that this formulation can support flexible magnetic control policies trained in simulation and deployed on experimental devices \citep{deepmind,practicalRLtokamak}. In the present work, the rigid RZIP SSM provides a fast physics-informed transition model for offline learning, while the learned policy maps real-time tracking errors of $I_p$, $R_c$, and $Z_c$ to voltage commands for the CS and PF coils.
\subsubsection{Decision Process Formulation}
\label{sec:decision_process_formulation}
Based on the above formulation, the EXL-50U magnetic-control task is defined by its observation space, action space, transition model, and reward function. Because the full RZIP state includes coil-current dynamics that are not directly provided to the policy, the deployed controller operates under partial observability and uses measured or reconstructed plasma quantities as observations. The observation and action spaces are summarized in table~\ref{tab:rl_magnet_control}.

\begin{table}
  \centering
  \begin{tabular}{lc}
    Component & Description \\[5pt]
    Observation Space $\mathcal{O}$ & 3-dimensional measurement vector ($I_p$, $R_c$, $Z_c$) \\[3pt]
    Action Space $\mathcal{A}$ & 6-dimensional vector (voltage commands for CS + 5 PF pairs)
  \end{tabular}
  \caption{Observation and action space definition for RL-based plasma control}
  \label{tab:rl_magnet_control}
\end{table}

\paragraph{Observation}
At each time step, the environment provides real-time measurements of $I_p^{\text{obs}}$, $R_c^{\text{obs}}$, and $Z_c^{\text{obs}}$. Before entering the policy network, these measurements are converted into tracking errors with respect to the reference trajectories:
\begin{equation}
\boldsymbol{o_t} = \begin{bmatrix}
e_{I_p}(t) , e_R(t) , e_Z(t)
\end{bmatrix}^T
\end{equation}
where $e_{I_p} = I_p^{\text{ref}} - I_p^{\text{obs}}$, $e_R = R_c^{\text{ref}} - R_c^{\text{obs}}$, and $e_Z = Z_c^{\text{ref}} - Z_c^{\text{obs}}$. This preprocessing aligns the policy input with the tracking objective.
\paragraph{Action Space}
Considering the vertical symmetry of the EXL-50U PF coil system, the current waveforms of vertically symmetric coil pairs are nearly overlapping during actual discharges. We therefore group PF1-10 into pairs and apply identical voltage commands to each pair throughout training and deployment, which reduces the control dimensionality while remaining consistent with the physical discharge characteristics. As the RL policy inference time exceeds the fast control cycle required for vertical stability, the VS coils are excluded from the action space during training and are governed by the conventional PID controller in real-time deployment.
The agent outputs a normalized 6-dimensional action vector $\boldsymbol{a_t} \in [-1,1]^6$, corresponding to the CS coil and five PF coil pairs: PF1--2, PF3--4, PF5--6, PF7--8, PF9--10. Actions are linearly mapped to the physical voltage limits of the corresponding coil power supplies via:
\begin{equation}
V_t^{(i)} = V_{\text{min}}^{(i)} + \frac{1 + a_t^{(i)}}{2}
\left( V_{\text{max}}^{(i)} - V_{\text{min}}^{(i)} \right)
\end{equation}
where $i$ denotes the index of the control channel ($i=1$ for CS, $i=2{\sim}6$ for the five PF pairs). For each PF coil pair, the mapped voltage $V_t^{(i)}$ is applied to both coils in the pair simultaneously. $V_t$ also serves as the control input $\boldsymbol{u}$ to the RZIP model.
\paragraph{Reward Function Design}
The magnetic-control task contains several objectives that must be satisfied simultaneously. In this work, these objectives are the tracking of plasma current $I_p$, radial centroid position $R_c$, and vertical centroid position $Z_c$. At each control step, the error of each objective is first converted into a dimensionless scalar. Relative errors are used for $I_p$ and $R_c$, whose targets are non-zero and have different physical units, while an absolute error is used for $Z_c$ because its reference value is zero:
\begin{equation}
\epsilon_{I_p}=\left|\frac{I_p^{\mathrm{obs}}-I_p^{\mathrm{ref}}}{I_p^{\mathrm{ref}}}\right|,\quad
\epsilon_R=\left|\frac{R_c^{\mathrm{obs}}-R_c^{\mathrm{ref}}}{R_c^{\mathrm{ref}}}\right|,\quad
\epsilon_Z=\left|Z_c^{\mathrm{obs}}-Z_c^{\mathrm{ref}}\right|.
\end{equation}
Each error is then transformed into a bounded quality measure $q_i\in[0,1]$, where larger values indicate better tracking. Different nonlinear transforms are used because the controlled variables have different physical response characteristics. For $I_p$, a bounded SoftPlus-like logistic transform is used:
\begin{equation}
q_{I_p}(\epsilon)=
\mathrm{clip}\!\left(2\sigma\!\left[-\ln(19)\frac{\epsilon}{b_{I_p}}\right],0,1\right),
\end{equation}
where $\sigma(x)=1/(1+\exp(-x))$ and $b_{I_p}=0.15$. This mapping gives unit reward at zero error and then decays smoothly as the relative current error increases. It is used for $I_p$ because plasma-current evolution is inductive and comparatively slow, so large transient errors can occur during early training or after controller handover. A gradual SoftPlus-like decay avoids making such states almost unrewarded, preserving a useful learning signal while still favoring smaller current error.

For the position objectives, sigmoid transforms are used:
\begin{equation}
q_j(\epsilon)=\sigma\!\left[\ln(19)\left(1-2\frac{\epsilon}{b_j}\right)\right],
\quad j\in\{R,Z\},
\end{equation}
with $b_R=0.15$ and $b_Z=0.05$. Compared with the SoftPlus-like current reward, the sigmoid form behaves more like a soft tolerance function: it keeps a high quality score for small position errors but drops rapidly as the error approaches the prescribed scale. This is appropriate for centroid-position control, where deviations are more directly related to equilibrium displacement and stability margin. The smaller value of $b_Z$ reflects the stricter requirement on vertical displacement: a $5\,\mathrm{cm}$ vertical error is already regarded as poor-quality tracking, whereas $I_p$ and $R_c$ use a broader 15\% relative-error scale. These values are not final accuracy requirements; rather, they define the error ranges over which the reward changes most strongly, based on the expected transient errors and the physical sensitivity of each controlled variable.

The individual quality measures are combined into a single scalar step reward by an equally weighted average:
\begin{equation}
r_t=\frac{1}{3}\left(q_{I_p}+q_R+q_Z\right).
\end{equation}
The training objective maximizes the cumulative reward over the 100-step episode, thereby penalizing persistent tracking errors more strongly than isolated deviations. The reward framework also supports terminal penalties when safety or stability termination conditions are triggered; in the present limiter-control experiments, no additional action penalty was applied, so the learned policy was driven primarily by the accumulated tracking quality of $I_p$, $R_c$, and $Z_c$.

\subsubsection{Training Setup and Convergence}
\label{sec:training_setup}
Key training hyperparameters are summarized in table~\ref{tab:training_hyperparams}. The framework is Stable-Baselines 3 \citep{stable-baselines3}. The agent interacts with the RZIP SSM environment in 100-step episodes (corresponding to 100\,ms of real time), with a total of 500 training iterations ($1.024 \times 10^6$ total timesteps). Each iteration collects 2048 environment steps for policy and value function updates. Episodes are initialized with small perturbations to the plasma state to improve the robustness of the learned policy.

\begin{table}
  \centering  
  \begin{tabular}{l|c}  
    Parameter & Value \\[6pt]
    Algorithm & PPO \\[2pt]
    Framework & Stable-Baselines 3 \\[2pt]
    Training iterations & 500 \\[2pt]
    Rollout per iteration & 2048 env steps \\[2pt]
    Total timesteps & $1.024 \times 10^6$ \\[2pt]
    Episode length & 100 steps (100\,ms) \\[2pt]
    Actor network architecture & [32, 32] \\[2pt]
    Critic network architecture & [64, 32] \\[2pt]
    Activation function & Tanh \\[2pt]
    Learning rate & $3 \times 10^{-4}$ \\[2pt]
    Clip range & 0.2 \\[2pt]
  \end{tabular}
  \caption{Key hyperparameters for PPO policy training}  
  \label{tab:training_hyperparams}  
\end{table}

The computational efficiency of the PPO algorithm was quantified for practical deployment. The training procedure consists of three phases: initial exploration (0–200k steps), rapid learning (200k–500k steps), and fine-tuning (500k–1M steps). Using an NVIDIA RTX 4090 GPU, 1 million training steps were completed in 30–40 minutes, which is orders of magnitude faster than conventional manual PID tuning (typically days to weeks of on-site iterative adjustment).

\begin{figure}
    \centering
    \includegraphics[width=0.75\linewidth]{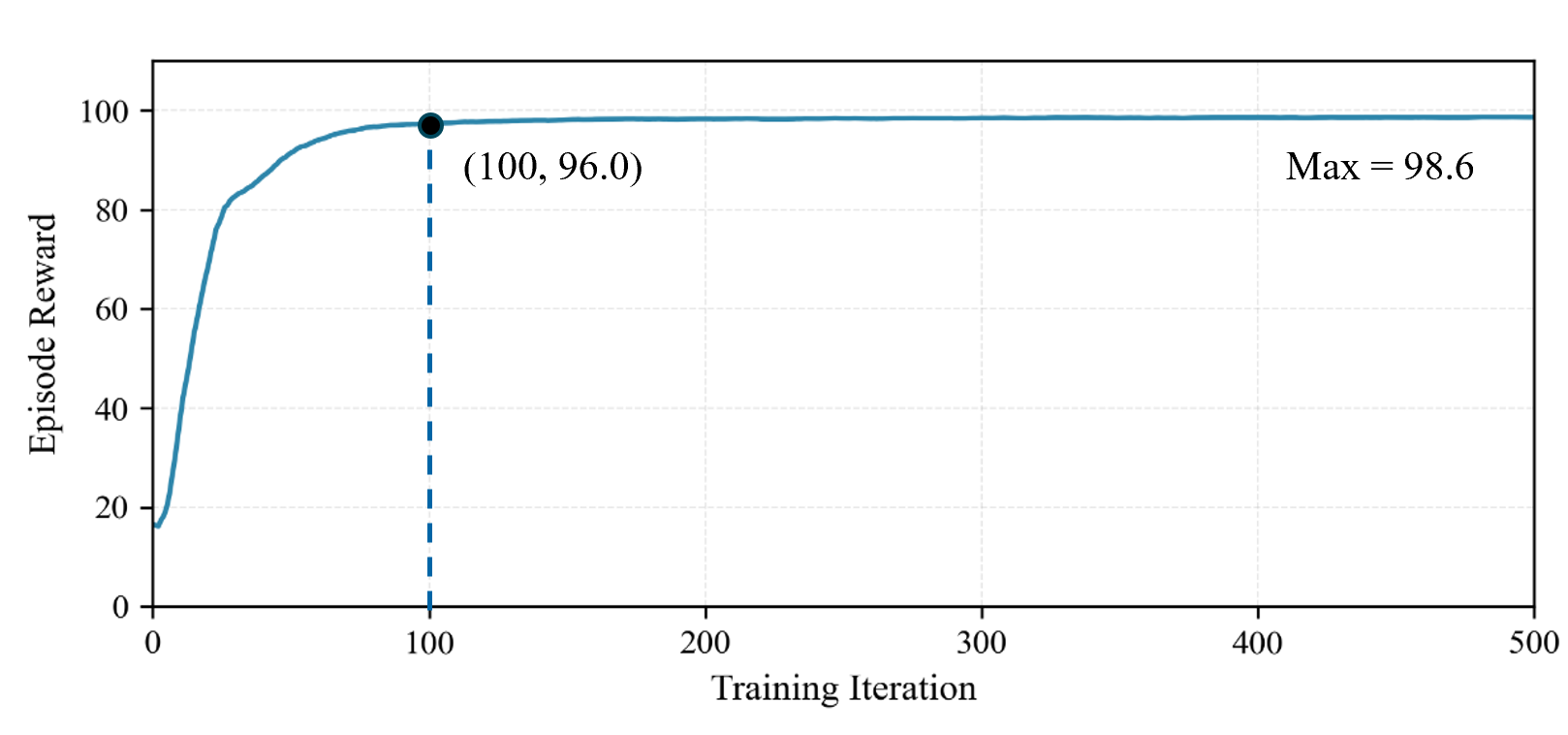}
    \caption{Training convergence curve of the PPO policy: mean episode reward versus training iteration.}
    \label{fig:training_curve}
\end{figure}

The training convergence curve is shown in figure~\ref{fig:training_curve}. The mean episode reward rises rapidly in the early training phase, reaching ~96 at 100 iterations, before entering a fine-tuning stage and stabilizing at 98.60 after 500 iterations (98.57 ± 0.04 over the final 50 iterations), confirming stable and consistent policy optimization in the SSM environment.

\subsubsection{Closed-Loop Simulation Validation}
\label{sec:sim_validation}
Prior to hardware deployment, the trained policy was validated in extended closed-loop simulations to assess its temporal generalization capability and robustness to real-world perturbations. This step is critical to eliminate obvious policy errors and ensure reliable sim-to-real transfer.

\begin{figure}
    \centering
    \includegraphics[width=0.85\linewidth]{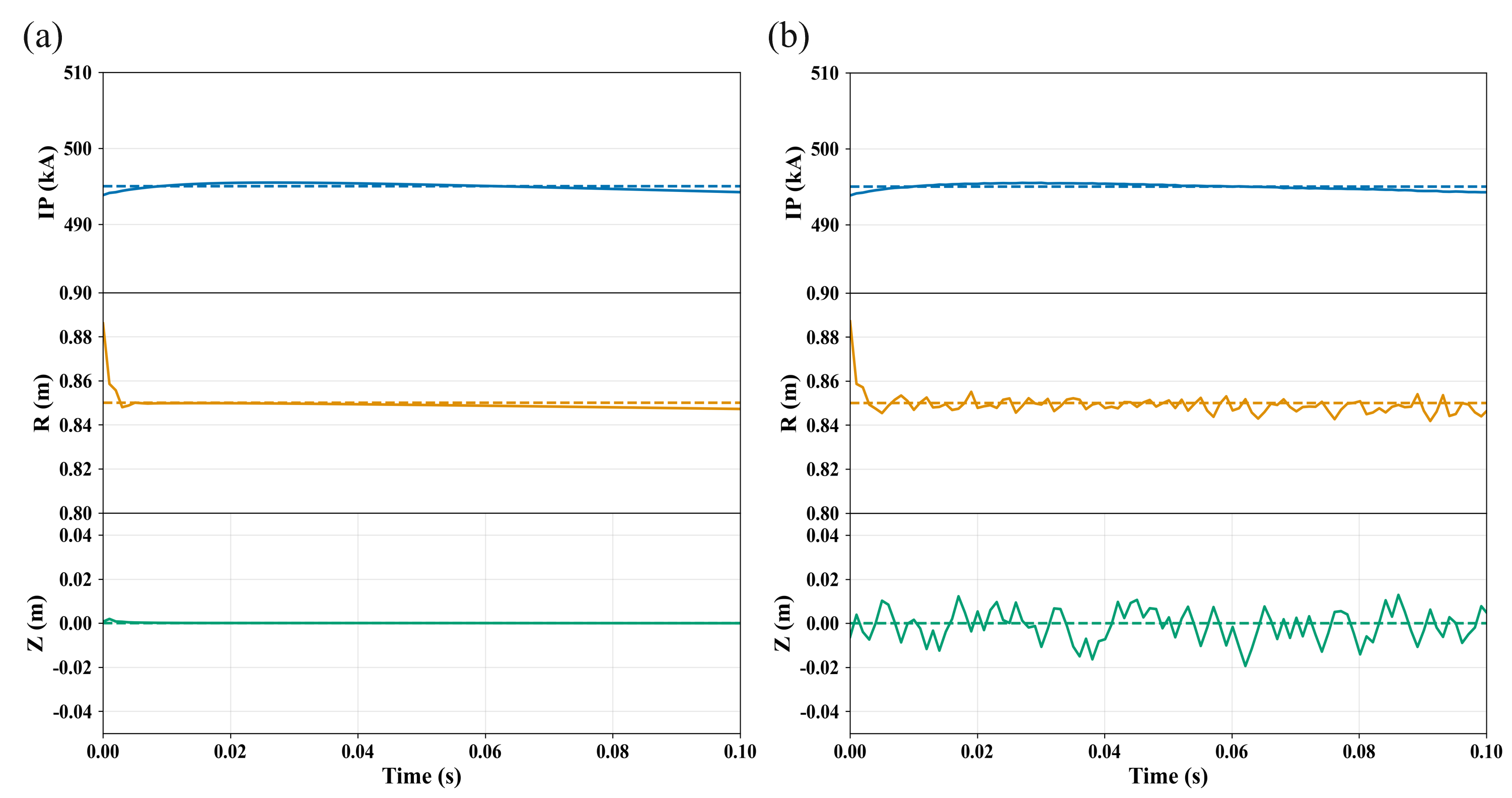}
    \caption{Closed-loop simulation results in the RZIP SSM environment. (a) Baseline control performance under nominal conditions. (b) Robustness test with measurement noise.}
    \label{fig:sim_closedloop}
\end{figure}

Under nominal conditions (500\,ms simulation, target $I_p=495$\,kA, $R_c=0.85$\,m, $Z_c=0$\,m), the controller reaches steady state after an initial ~5\,ms transient, regulating $R_c$ to within ±5\,mm, $Z_c$ to within ±1\,mm, and $I_p$ to within 2\,kA of the target (figure~\ref{fig:sim_closedloop}(a)). When realistic Gaussian measurement noise ($\sigma_{I_p}=6$\,kA, $\sigma_{R_c,Z_c}=2$\,cm) and 2\,ms feedback latency are introduced, the policy maintains stable control of all quantities within safe operating limits, with only moderate oscillations in the vertical position (figure~\ref{fig:sim_closedloop}(b)). These results confirm the policy’s robustness and suitability for hardware deployment on the EXL-50U device.

\subsection{Real-Time Deployment on EXL-50U PCS}
\label{sec:deployment}
\subsubsection{Policy Inference Integration}
The trained RL policy is a neural network implemented in PyTorch. For real-time inference, the PyTorch policy network was optimized with NVIDIA TensorRT, achieving an average inference latency of 0.056,ms, a 99th percentile latency of 0.064,ms, and a maximum latency of 0.413,ms. These values are well within the 1\,ms main control cycle requirement of the EXL-50U PCS, confirming the practical feasibility of the proposed RL control scheme. The optimized policy is integrated into the PCS as a callable function that receives a 3-dimensional state vector and outputs a 6-dimensional action vector.

\begin{figure}
    \centering
    \includegraphics[width=0.8\linewidth]{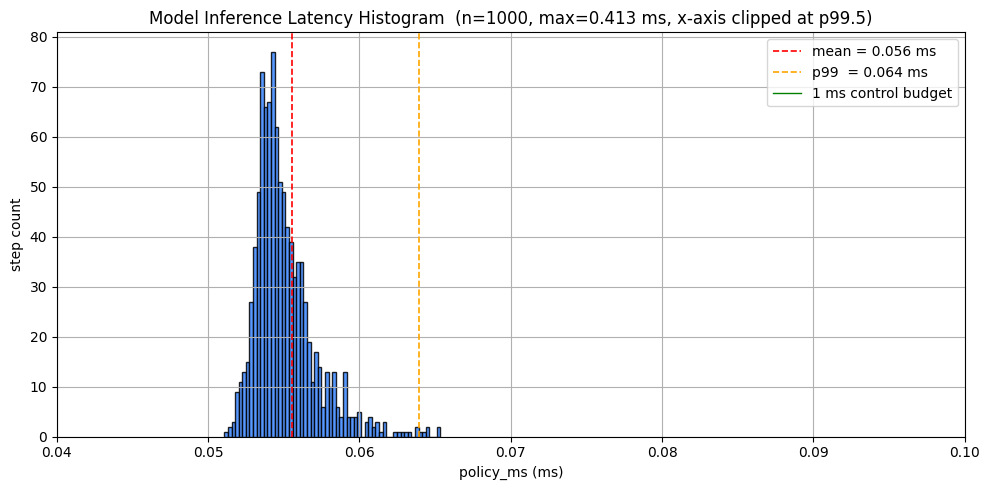}
    \caption{Inference latency distribution of the optimized RL policy.}
    \label{fig:Inference_latency}
\end{figure}

\subsubsection{Online Control Workflow}

The online workflow follows the sequence illustrated in figure~\ref{fig:framework}, with the following steps executed every control cycle.

\textbf{Step 1: Diagnostic data acquisition and state estimation.} Real-time magnetic signals are acquired via the reflective memory (RFM) network. The plasma current \(I_p\) is directly measured by a Rogowski coil. The centroid position \((R_c, Z_c)\) is reconstructed from four symmetric tangential magnetic probes using the RZEstimator on EXL-50U:
\begin{align}
Z_c &= \frac{1}{C_Z I_p} \left( -B_{\rm LL} + B_{\rm LU} + B_{\rm RU} - B_{\rm RL} \right) \nonumber \\
R_c - R_0 &= \frac{1}{C_R I_p} \left( -B_{\rm LL} - B_{\rm LU} + B_{\rm RU} + B_{\rm RL} \right) \nonumber
\end{align}
where \(B_{\mathrm{LU}}, B_{\mathrm{LL}}, B_{\mathrm{RU}}, B_{\mathrm{RL}}\) are the magnetic signals, \(R_0\) the reference major radius, and \(C_Z, C_R\) calibration constants.

The errors are then computed as the differences between the observed values and the reference trajectories:
\[
e_{I_p} = I_p^{\mathrm{ref}} - I_p^{\mathrm{obs}},\quad
e_{R} = R_c^{\mathrm{ref}} - R_c^{\mathrm{obs}},\quad
e_{Z} = Z_c^{\mathrm{ref}} - Z_c^{\mathrm{obs}}.
\]
These three errors form the 3-dimensional state vector.

\textbf{Step 2: Mode selection.} The PCS can operate in either conventional PID mode or RL mode. The selection is made prior to the discharge and remains fixed during the flattop phase; a soft transition period is used to smoothly hand over control from PID to RL when RL mode is activated.

\begin{itemize}
\item \textbf{PID mode:} The errors are fed into independent incremental PID controllers. The PID outputs are current adjustments for each coil. These are converted to voltage adjustments via a mutual inductance matrix and then added to pre-computed feedforward voltages. The resulting voltage commands are sent to the power supplies.

\item \textbf{RL mode:} The 3-dimensional error vector is directly fed into the pre-trained TensorRT-optimized policy network. The network outputs 6-dimensional normalized actions, which are linearly mapped to physical voltage limits for the CS coil and five symmetric PF coil pairs. These voltage commands bypass the PID and feedforward modules and are sent directly to the power supplies.
\end{itemize}

\textbf{Step 3: Actuation.} The voltage commands are output to the coil power supplies via the RFM network, completing the closed-loop control within the 1~ms cycle.

\section{Experimental Results}
In this section, we present the results of dedicated RL-based control experiments conducted on the EXL-50U ST. The RL controller was trained offline within the rigid RZIP SSM initialized to the equilibrium state of reference limiter discharge \#12030 at 500\,ms, then deployed online to take over control of the CS and PF1–PF10 during the flattop phase of plasma discharges. A 10\,ms soft transition period was implemented to ensure smooth handover between the conventional PID controller and the RL-based controller. Over 10 repeated discharges were conducted under both limiter and divertor configurations to comprehensively evaluate control performance, robustness, and cross-configuration generalization capability.

\subsection{Limiter Configuration Control Performance}
\label{sec:limiter_performance}
The baseline control scheme adopts independent incremental PID loops for each coil to track the target plasma current $I_p$ and centroid position $(R_c, Z_c)$. In contrast, the RL-based controller directly outputs synchronized voltage commands to all CS and PF coils to achieve the same control objectives. Owing to latency constraints, vertical displacement control remains implemented via the conventional fast PID loop in all experiments.

\begin{figure}
    \centering
    \includegraphics[width=0.75\linewidth]{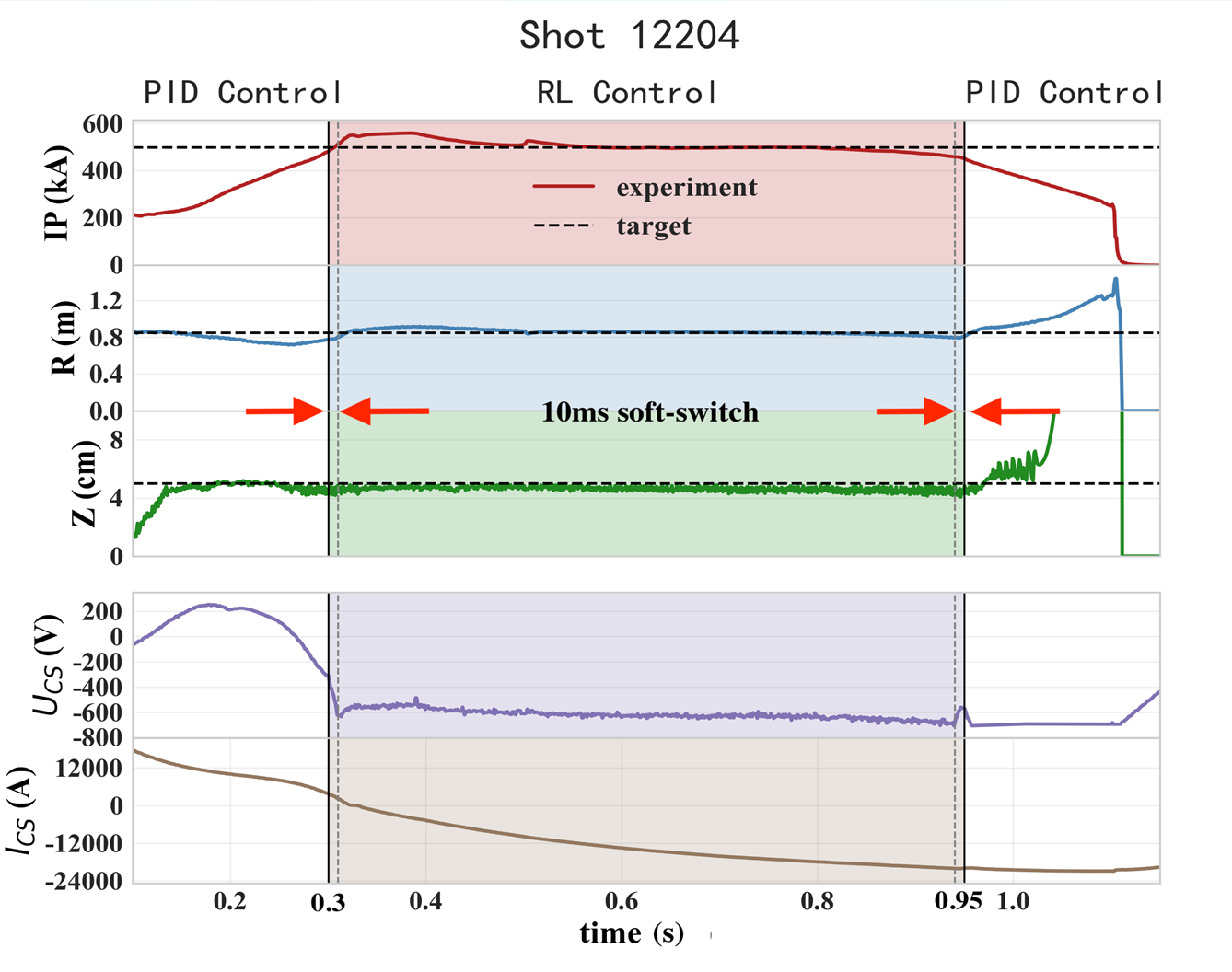}
    \caption{Key waveforms of limiter discharge \#12204 under RL-based control. Shown are plasma current $I_p$, radial centroid position $R_c$, and vertical centroid position $Z_c$ over time. Horizontal black dashed lines denote target values. The shaded region (300–950\,ms) indicates the RL control phase.}
    \label{fig:limiter_12204_discharge}
\end{figure}

Figure \ref{fig:limiter_12204_discharge} presents the key waveforms of limiter discharge \#12204 under RL control. The RL controller took over at 300\,ms via the 10\,ms soft transition and maintained stable operation until 950\,ms, achieving a continuous 650\,ms of RL-controlled limiter discharge. During the RL controller takeover period, the average deviation of the plasma current was approximately 6 kA.  

\subsection{Performance Comparison with Conventional PID Control}
\label{sec:rl_vs_pid}

\begin{figure}
    \centering
    \includegraphics[width=0.75\linewidth]{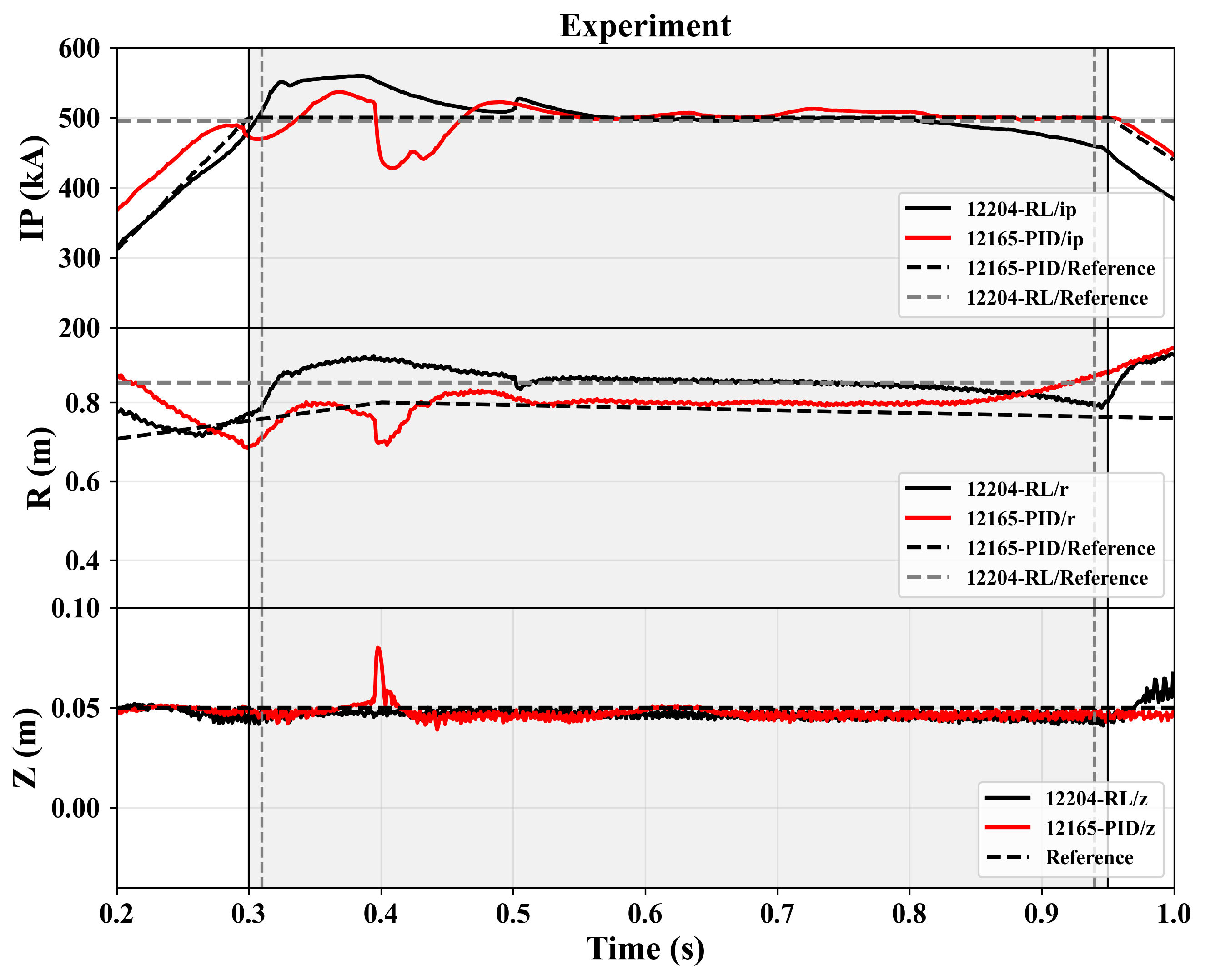}
    \caption{Comparison of key plasma parameter waveforms between RL-controlled discharge \#12204 (black lines) and PID-controlled baseline discharge \#12165 (red lines) under identical limiter configuration. Shown are $I_p$, $R_c$ and $Z_c$ over time. Dashed lines denote the respective target reference values for each parameter. The shaded region (300–950\,ms) indicates the RL control phase.}
    \label{fig:limiter_rl_vs_pid}
\end{figure}

To comprehensively evaluate the RL controller’s performance against the baseline scheme, we compare two limiter discharges under similar target parameters ($I_p = 500$ kA, $R_c \approx 0.85$ m, $Z_c = 0.05$ m) and identical operating conditions: the RL-controlled discharge \#12204 and the PID-controlled baseline discharge \#12165. Figure~\ref{fig:limiter_rl_vs_pid} presents the key plasma parameter waveforms for both discharges, while table~\ref{tab:limiter_performance_metrics} summarizes the quantitative performance metrics.

\begin{table}
    \centering
    \setlength{\tabcolsep}{6pt}
    \begin{tabular}{lccccc}
        Metric & Parameter & Unit & PID (\#12165) & RL (\#12204) \\[10pt]
        Mean Squared Error (MSE) & $I_p$ & kA² & 54.0 & 89.0 \\
        & $R_c$ & m² & $2.74 \times 10^{-4}$ & $4.79 \times 10^{-5}$ \\[6pt]
        Max/Mean Offset & $I_p$ & kA & 19.37 / 5.79 & 32.08 / 5.89 \\
        & $R_c$ & m & 0.028 / 0.016 & 0.017 / 0.005 \\[6pt]
        Standard Deviation (STD) & $I_p$ & kA & 4.53 & 7.37 \\
        & $R_c$ & m & 0.0052 & 0.0036 \\
    \end{tabular}
       \caption{Quantitative performance metrics of RL and PID control under limiter configuration. All metrics are computed over the stable flattop phase (500–800\,ms) to exclude transient effects from the RL controller takeover.}
    \label{tab:limiter_performance_metrics}
\end{table}

From the waveforms in figure~\ref{fig:limiter_rl_vs_pid}, the PID-controlled discharge exhibits significant transient deviations in $I_p$ and $R_c$ during the early phase (300--400\,ms), whereas the RL-controlled discharge shows a smoother transient response with rapid $I_p$ rise and steady $R_c$ convergence. Both schemes achieve stable $Z_c$ tracking via the dedicated fast PID loop.

Quantitatively, RL outperforms PID in radial position tracking: the MSE of $R_c$ is reduced by 41\% (from $1.49 \times 10^{-3}$ to $8.82 \times 10^{-4}$~m$^2$), the standard deviation by 31\% (from 0.0052 to 0.0036~m), and the mean offset by 69\% (from 0.016 to 0.005~m).

For plasma current control, the average tracking error is comparable between the two schemes . However, RL exhibits a slightly higher standard deviation and maximum offset during the stable phase, likely due to the RL policy's multi-objective optimization that prioritizes radial position regulation.

\subsection{Key Findings and Control Strategy Analysis}
\label{sec:control_strategy_analysis}
\begin{figure}
    \centering
    \begin{subfigure}[b]{0.48\textwidth}
        \centering
        \includegraphics[width=\textwidth]{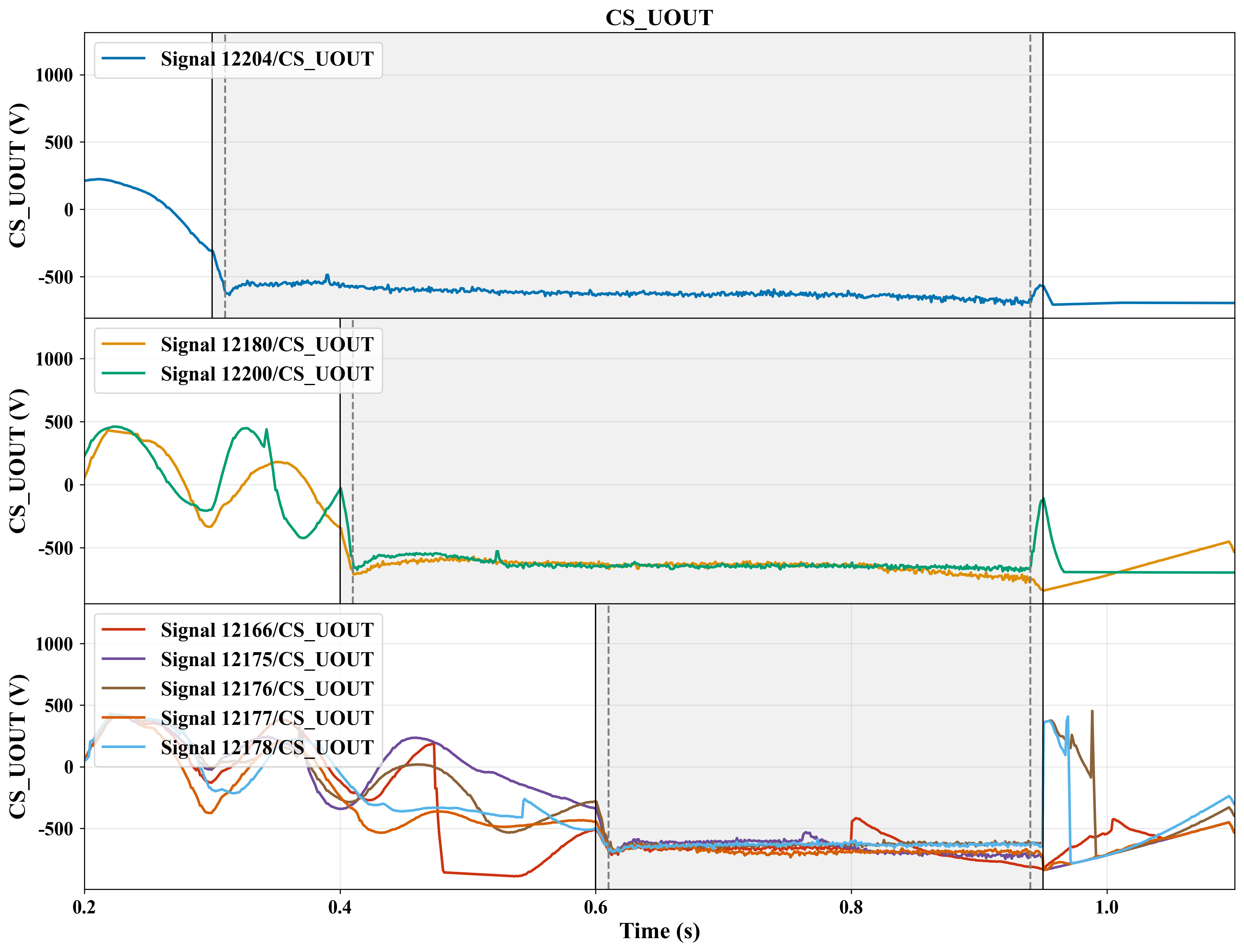}
        \caption{CS coil voltage commands for multiple limiter discharges.}
        \label{fig:CS_UOUT}
    \end{subfigure}
    \hfill
    \begin{subfigure}[b]{0.48\textwidth}
        \centering
        \includegraphics[width=\textwidth]{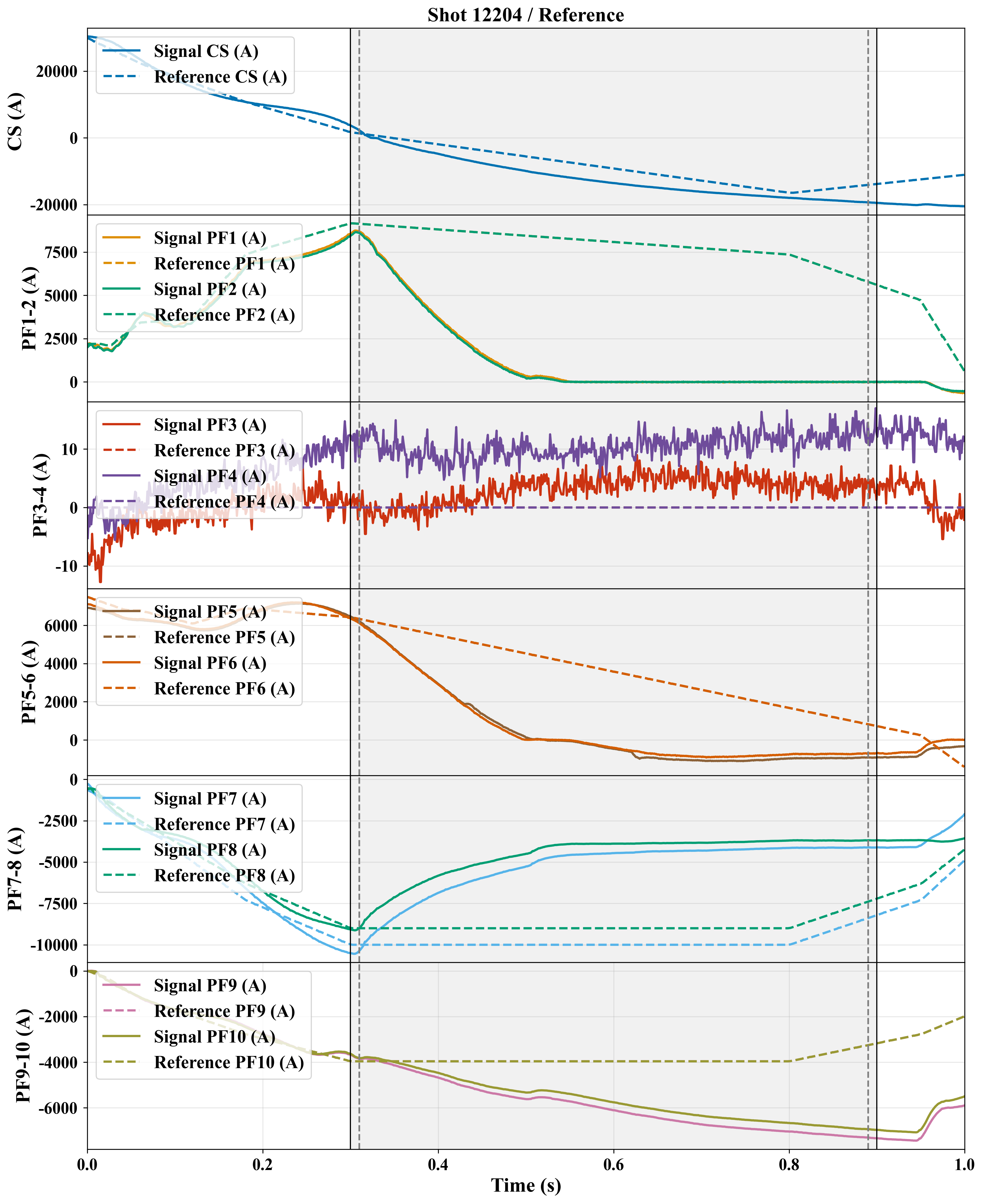}
        \caption{PF coil currents for limiter discharge \#12204. Dashed lines denote PID feedforward references.}
        \label{fig:PF_currents}
    \end{subfigure}
    \caption{Coil control signals under RL-based magnetic control. (a) CS voltage commands stabilize to near-constant values after RL activation. (b) RL-induced deviations from PID feedforward references, showing optimal redistribution of control effort across actuators.}
    \label{fig:coil_waveforms}
\end{figure}

Beyond basic tracking performance, the RL controller exhibits a fundamentally different control strategy compared to the PID scheme, as revealed by the coil control signals in figure~\ref{fig:coil_waveforms}.

As shown in figure~\ref{fig:coil_waveforms}(a), upon handover to RL control, the CS voltage commands rapidly stabilize to near-constant values with only minor fluctuations, in stark contrast to the large, high-frequency adjustments required by PID control. This behavior reveals that the RL policy identifies a quasi-equilibrium operating point for the CS coil that sustains the target $I_p$ without active modulation, effectively offloading fine-grained $I_p$ regulation to the PF coils.

Correspondingly, the PF coil current waveforms in figure~\ref{fig:coil_waveforms}(b) illustrate how the RL policy optimally redistributes control authority across actuators to achieve this offloading. Specifically, PF9 and PF10 currents exceed the PID feedforward references, while PF7 and PF8 currents are accordingly reduced. This redistribution reflects the policy’s learned understanding of actuator effectiveness: PF9/10 exhibit stronger radial position control authority than PF7/8, so the RL controller preferentially leverages these more efficient coils. Additionally, PF1/2 and PF5/6 currents are driven to near-zero levels. This is a direct consequence of the training objective, which prioritizes only $I_p$, $R_c$, and $Z_c$ regulation, leaving coils dedicated to plasma shaping not actively engaged.

Taken together, these observations confirm that RL can automatically exploit the inherent electromagnetic coupling between coils to discover novel equilibrium control solutions. Achieving this with conventional PID controllers is difficult without extensive, physics-informed manual tuning of cross-channel interactions.

\section{Discussion and Conclusion}
\label{sec:discussion_conclusion}

The experimental results presented above demonstrate that the proposed SSM-enabled RL framework achieves stable regulation of plasma current and centroid position in EXL-50U limiter discharges. Most importantly, the RL controller discovers a more efficient control strategy than conventional PID, leveraging inter-coil coupling to redistribute control authority and reduce unnecessary high-frequency actuation. However, the current framework still has several notable limitations that constrain its broader application to high-performance spherical torus operations.

\subsection{Limitations of the Current RL Control Framework}
First, the PF9/10-dominant, PF7/8-suppressed current distribution identified in Section~\ref{sec:control_strategy_analysis} may compromise vertical stability margins under high-performance plasma conditions, where VDEs grow at significantly higher rates. This underscores the critical need to incorporate stability-aware objectives into the RL reward function.

Second, the policy exhibits limited cross-configuration generalization, as pre-trained limiter policies cannot fully sustain stable divertor operation. This is a fundamental limitation of the linear RZIP SSM used for training, which only captures global plasma current and position dynamics, neglecting shape deformation and X-point physics critical for divertor equilibrium.

Third, the training environment is idealized and does not incorporate realistic disturbances such as MHD instabilities, impurity influx, or auxiliary heating perturbations, which may degrade controller performance under more complex experimental conditions. Additionally, the current policy is trained as an unconstrained optimization problem, lacking explicit hard safety guarantees for coil current/voltage limits and plasma operational boundaries, requiring a fallback PID controller and interlock system for device protection.

Finally, the black-box nature of neural network controllers hinders interpretability of control decisions, complicating engineering debugging and systematic performance optimization.

\subsection{Future Research Directions}
Addressing these limitations motivates the following targeted future research directions, aligned with the long-term control goals of EXL-50U and the next-generation EHL-2 spherical torus. EHL-2 aims to verify the thermal reaction rates of $\mathrm{p}^{-11}\mathrm{B}$ fusion, establish spherical torus experimental scaling laws at $10\,\mathrm{keV}$ ion temperature, and provide a design basis for subsequent experiments to test and realize $\mathrm{p}^{-11}\mathrm{B}$ fusion burning plasma \citep{liang2025overview,syj}. In particular, extending the RL framework to multi-objective plasma shape control is a high priority: by incorporating elongation, triangularity, and X-point position into the reward function and training environment, the controller can be upgraded to support full equilibrium control of divertor configurations, which is essential for high-confinement plasma operation.

Additionally, integrating hard operational constraints (e.g., VDE growth rate, edge safety factor, and coil hardware limits) into the RL algorithm via constrained policy optimization will enhance controller safety and robustness. Future work will also aim to extend the control horizon from the flattop phase to the full discharge cycle (breakdown, ramp-up, and ramp-down), and develop hybrid physics-data-driven training environments that combine the RZIP model with experimental data to improve cross-regime generalization. Finally, combining RL controllers with conventional PID architectures—where RL handles complex coupled dynamics and PID provides fast, reliable backup for safety-critical tasks—will create a practical, high-performance control system suitable for long-pulse operation.

\subsection{Conclusion}
In summary, this work demonstrates the feasibility of reinforcement learning for magnetic control of EXL-50U spherical torus plasmas. By leveraging a computationally efficient rigid RZIP SSM for offline training, the proposed RL controller achieves stable, low-latency regulation of plasma current and centroid position in limiter discharges, and discovers a novel control strategy that outperforms conventional PID in terms of actuation efficiency. While further improvements are needed in safety, generalization, and control scope, these results lay a critical foundation for the development of next-generation autonomous control systems for spherical torus fusion devices.

\section*{Acknowledgments}

This work was supported by the National Natural Science Foundation of China (Grant No 12435014). We would like to express our gratitude to Yuejiang Shi and Xianming Song, both chief scientists of EXL-50U. Dr. Shi provided crucial algorithm optimization suggestions during the experiment, which helped the RL‑control experiment succeed on the EXL‑50U platform, and Dr. Song designed the plasma control scheme for the RL‑control experiment. We are also grateful to Professor Bingjia Xiao of the Institute of Plasma Physics for his valuable guidance.

Only basic grammar and spelling checking tools were used in the preparation of this manuscript; no generative AI was employed to produce text, data, or figures.

\bibliographystyle{jpp}

\bibliography{jpp-instructions}

\end{document}